\documentclass[english]{revtex4-1}
\usepackage[T1]{fontenc}
\usepackage[latin9]{inputenc}
\usepackage{amsmath}
\usepackage{amssymb}
\usepackage{graphicx}
\usepackage{wasysym}

\makeatletter
\numberwithin{equation}{section}

\usepackage{babel}

\makeatother

\usepackage{babel}
\begin{document}

\title{Coherent states for ladder operators of general order related to
exceptional orthogonal polynomials}

\author{Scott E. Hoffmann\textsuperscript{1}, Véronique Hussin\textsuperscript{2},
Ian Marquette\textsuperscript{1} and Yao-Zhong Zhang\textsuperscript{1}}

\address{\textsuperscript{1}School of Mathematics and Physics, The University
of Queensland, Brisbane, QLD 4072, Australia~\\
~\\
\textsuperscript{2}Département de Mathématiques et de Statistique,
Université de Montréal, Montréal, Québec, H3C 3J7, Canada}
\email{scott.hoffmann@uqconnect.edu.au}

\begin{abstract}
We construct the coherent states of general order, $m,$ for the ladder operators $c(m)$ and $c^{\dagger}(m)$, which act
on rational deformations of the harmonic oscillator. The position
wavefunctions of the eigenvectors involve type III Hermite exceptional
orthogonal polynomials. We plot energy expectations, time-dependent
position probability densities for the coherent states and for the
even and odd cat states, Wigner functions, and Heisenberg uncertainty
relations. We find generally non-classical behaviour, with one exception:
there is a regime of large magnitude of the coherent state parameter,
$z,$ where the otherwise indistinct position probability density
separates into $m+1$ distinct wavepackets oscillating and colliding
in the potential, forming interference fringes when they collide.
The Mandel $Q$ parameter is calculated to find sub-Poissonian statistics,
another indicator of non-classical behaviour. We plot the position
standard deviation and find squeezing in many of the cases. We calculate
the two-photon-number probability density for the output state
when the $m=4,\mu=-5$ coherent states (where $\mu$ labels the lowest
weight in the superposition) are placed on one arm of a beamsplitter.
We find that it does not factorize, again indicating non-classical
behaviour. Calculation of the linear entropy for this beamsplitter
output state shows significant entanglement, another non-classical
feature. We also construct linearized versions, $\tilde{c}(m),$ of
the annihilation operators and their coherent states and calculate
the same properties that we investigate for the $c(m)$ coherent states.
For these we find similar behaviour to the $c(m)$ coherent states,
at much smaller magnitudes of $z,$ but comparable average energies.
\end{abstract}
\maketitle

\section{Introduction}

In a recent publication \cite{Hoffmann2018a}, the present authors
constructed the coherent states associated with the ladder operators
$c(m)$ and $c^{\dagger}(m)$ \cite{Marquette2013b,Marquette2013a}
for the particular choice of the order $m=2,$ and investigated their
properties. In this paper we construct the coherent states for general
order, $m$ (even). The systems are rational, nonsingular, deformations
of the harmonic oscillator, obtained through supersymmetric quantum
mechanics (SUSY QM), with position wavefunctions involving type III
Hermite exceptional orthogonal polynomials \cite{GomezUllate2010,Marquette2013b,Marquette2013a}.
The ladder operators $c(m)$ and $c^{\dagger}(m)$ are distinguished
in that they are currently the only ones that connect all states of
the partner Hamiltonians, including the ground states.

A coherent state vector is a coherent superposition of the energy
eigenvectors of a system. Coherent states are of widespread interest
\cite{Glauber1963a,Glauber1963b,Klauder1963a,Klauder1963b,Barut1971,Perelomov1986,Gazeau1999,Quesne2001,Fernandez1995,Fernandez1999,GomezUllate2014,Ali2014}
because they can, in some cases, be the quantum-mechanical states
with the most classical behaviour.

Since the ladder operators $c(m)$ and $c^{\dagger}(m)$ change the
energy index by $m+1,$ there are $m+1$ distinct, orthogonal coherent
states, each with a different lowest weight state. For these ladder
operators, we construct the coherent states (defined as eigenvectors
of the annihilation operator $c(m)$ with complex eigenvalue, $z$).
Then we calculate energy expectations, time-dependent position probability
densities for the coherent states and for the even and odd cat states,
Wigner functions and Heisenberg uncertainty relations. The Mandel
$Q$ parameter is calculated to find the number statistics of these
coherent states. The position standard deviation is plotted as a function
of $\mathrm{Re}\,z$ and $\mathrm{Im}\,z$ to investigate the possibility
of squeezing. We place one of our coherent states on one arm of a
beamsplitter and calculate the two-photon-number probability distribution
to look for factorization. The linear entropy is calculated for this
output state to look for entanglement.

For comparison, we also construct linearized versions of the ladder
operators, $\tilde{c}(m)$ and $\tilde{c}^{\dagger}(m),$ and the
coherent states of $\tilde{c}(m),$ investigate the same properties
and compare with the $c(m)$ coherent states.

In the previous paper \cite{Hoffmann2018a}, we compared coherent
states labelled $\mathrm{CS}(c(2),\mu)$ and $\mathrm{CS}(\tilde{c}(2),\mu)$
(for lowest weights $\mu=-3,1,2$) at the same value of the complex
coherent state parameter, $z,$ and found significantly different
behaviours. However the average energies of these coherent states
have significantly different dependencies on $|z|$. In this paper,
we will compare the two cases, $\mathrm{CS}(c(m),\mu)$ and $\mathrm{CS}(\tilde{c}(m),\mu),$
at comparable average energies, and we will find similar rather than
different behaviours.

We have used the Barut-Girardello definition \cite{Barut1971} of
coherent states as eigenvectors of the annihilation operator with
complex eigenvalue, $z.$ Note that there are other possible definitions,
for example using a displacement operator \cite{Perelomov1986}.

Coherent states for various choices of ladder operators in systems
that are the rational extensions of the harmonic oscillator were also
constructed in \cite{Bermudez2015}.

In Section II we review the ladder operators $c(m)$ and $c^{\dagger}(m)$
and their matrix elements. In Section III we form the coherent states
$\mathrm{CS}(c(m),\mu)$, labelled by the lowest weight, $\mu,$ in
each superposition. In Section IV we calculate energy expectations
for $\mathrm{CS}(c(m),\mu)$ and $\mathrm{CS}(\tilde{c}(m),\mu)$
for all cases of $m=4$ and $m=6.$ In Section V we plot the time-dependent
position probability densities for several cases. In Section VI we
plot time-dependent position probability densities for the even and
odd Schrödinger cat states formed from $\mathrm{CS}(c(6),-7)$ at
$z=10^{8}.$ In Section VII we plot the Wigner functions for all cases
of $\mathrm{CS}(c(6),\mu)$ and $z=10$. In Section VIII we calculate
Heisenberg uncertainty relations and find an example of squeezing.
In Section IX we calculate the Mandel parameter for $\mathrm{CS}(c(4),-5)$
and find sub-Poissonian number statistics. In Section X we place the
coherent state $\mathrm{CS}(c(4),-5)$ on one arm of a beamsplitter,
show that the output two-quantum number distribution does not factorize
and calculate the linear entropy of the output state. Conclusions
follow in Section XI.

\section{The ladder operators, $c(m),$ and their matrix elements for general
$m$}

The systems we consider are rational deformations of the harmonic
oscillator, obtained through supersymmetry \cite{Marquette2013b,Marquette2013a},
with Hamiltonians $H^{+}=-\frac{d^{2}}{dx^{2}}+x^{2}+2m+1$ and $H^{-}=-\frac{d^{2}}{dx^{2}}+V^{(m)}(x)$,
with the dependence on $m$ of $H^{+}$ and $H^{-}$ understood. The
partner potentials are, for any even $m,$
\begin{equation}
V^{(m)}(x)=x^{2}-2[\frac{\mathcal{H}_{m}^{\prime\prime}}{\mathcal{H}_{m}}-(\frac{\mathcal{H}_{m}^{\prime}}{\mathcal{H}_{m}})^{2}+1].\label{eq:2.1}
\end{equation}
The modified Hermite polynomials are
\begin{equation}
\mathcal{H}_{m}(x)\equiv(-i)^{m}H_{m}(ix),\label{eq:2.2}
\end{equation}
which are positive definite for even $m.$ The spectra of the partner
Hamiltonians and the harmonic oscillator Hamiltonians are the same
except for an additional ground state of the partner Hamiltonians
\begin{align}
E_{\nu}^{(+,m)} & =2(\nu+m+1)\quad\mathrm{for}\ \nu=0,1,2,\dots\ \mathrm{and}\ E_{\nu}^{(-,m)}=2(\nu+m+1)\quad\mathrm{for}\ \nu=-m-1,0,1,2,\dots,\label{eq:2.3}
\end{align}
with
\begin{equation}
H^{+}|\,\nu\,(+)\,\rangle=E_{\nu}^{(+,m)}|\,\nu\,(+)\,\rangle\quad\mathrm{for}\ \nu=0,1,2,\dots\ \mathrm{and}\ H^{-}|\,\nu\,(-)\,\rangle=E_{\nu}^{(-,m)}|\,\nu\,(-)\,\rangle\quad\mathrm{for}\ \nu=-m-1,0,1,2,\dots.\label{eq:2.3.1}
\end{equation}

The ladder operators, $c(m),$ are constructed and their matrix elements
derived in \cite{Marquette2013b}. Each operator involves a path from
$H^{-}$ to $H^{+}$ through the supercharge $A^{\dagger}$ then a
return to $H^{-}$ in $m$ steps through intermediate Hamiltonians
that may be singular. This is
\begin{equation}
c(m)=A_{m}\dots A_{2}A_{1}A^{\dagger}.\label{eq:2.4}
\end{equation}

We make a small change to the matrix elements compared to \cite{Marquette2013b}
to simplify subsequent calculations. We rephase the ground state,
$|\,-m-1\,(-)\,\rangle,$ multiplying it by $-1.$ This does not change
the polynomial Heisenberg algebra, which only involves diagonal operators.
Then we multiply all the position wavefunctions by $-1,$ which has
no physical consequence. Then we find the matrix elements
\begin{equation}
\langle\,\nu-m-1\,(-)\,|\,c(m)\,|\,\nu\,(-)\,\rangle=-[2^{m+1}(\nu-1)(\nu-2)\dots(\nu-m)(\nu+m+1)]^{\frac{1}{2}}\equiv a_{\nu}^{(m)}\label{eq:2.5}
\end{equation}
for all $\nu=-m-1,0,1,2,\dots.$ The operator $c(m)$ lowers the index
by $m+1.$

The polynomial Heisenberg algebra satisfied by these ladder operators
is \cite{Marquette2013b}
\begin{align}
[H^{-},c^{\dagger}] & =(2m+2)c^{\dagger},\quad[H^{-},c]=-(2m+2)c,\nonumber \\{}
[c,c^{\dagger}] & =Q_{m}(H^{-}+2m+2)-Q_{m}(H^{-}),\label{eq:2.5.1}
\end{align}
with
\begin{equation}
Q_{m}(H^{-}+2m+2)=(H^{-}+2m+2)\prod_{i=1}^{m}(H^{-}-2i)\quad\mathrm{and}\quad Q_{m}(H^{-})=H^{-}\prod_{i=1}^{m}(H^{-}-2m-2-2i),\label{eq:2.5.2}
\end{equation}
polynomials of order $m+1$ in $H^{-}.$

The position wavefunctions are
\begin{equation}
\psi_{\nu}^{(m)}(x)=\mathcal{N}_{\nu}^{(m)}\frac{e^{-x^{2}/2}}{\mathcal{H}_{m}(x)}y_{\nu+m+1}^{(m)}(x),\label{eq:2.6}
\end{equation}
with the Hermite exceptional orthogonal polynomials
\begin{align}
y_{0}^{(m)}(x) & =1,\nonumber \\
y_{\nu+m+1}^{(m)}(x) & =\mathcal{H}_{m}(x)H_{\nu+1}(x)+m\mathcal{H}_{m-1}(x)H_{\nu}(x),\label{eq:2.7}
\end{align}
and the normalization factors
\begin{align}
\mathcal{N}_{-m-1}^{(m)} & =\left[\frac{2^{m}m!}{\sqrt{\pi}}\right]^{\frac{1}{2}},\nonumber \\
\mathcal{N}_{\nu}^{(m)} & =\frac{1}{\sqrt{\sqrt{\pi}2^{\nu+1}(\nu+m+1)\nu!}}.\label{eq:2.8}
\end{align}

\section{The coherent states of the operators $c(m)$ and $\tilde{c}(m)$}

We have seen that the ladder operator $c(m)$ connects states with
indices differing by $m+1.$ This separates the space into $m+1$
distinct, orthogonal ladders, with lowest weights $\mu=-m-1,1,2,\dots,m.$
So there must be $m+1$ coherent states of $c(m),$ defined with the
Barut-Girardello definition \cite{Barut1971} as eigenstates of the
annihilation operator with complex eigenvalue $z,$ with each one
being a superposition of the state vectors of one of the $m+1$ ladders
of the form
\begin{equation}
|\,z,c,m,\mu\,\rangle=\sum_{k=0}^{\infty}|\,\mu+(m+1)k,m\,(-)\,\rangle A_{k}^{(m,\mu)}(z).\label{eq:3.1}
\end{equation}
Solving the defining condition,
\begin{equation}
c(m)\,|\,z,c,m,\mu\,\rangle=z\,|\,z,c,m,\mu\,\rangle,\label{eq:3.2}
\end{equation}
and normalizing gives
\begin{equation}
A_{k}^{(m,\mu)}(z)=\frac{1}{\sqrt{F^{(m,\mu)}(z)}}\frac{z^{k}}{D_{k}^{(m,\mu)}},\label{eq:3.3}
\end{equation}
with
\begin{equation}
D_{k}^{(m,\mu)}=\prod_{i=1}^{k}a_{\mu+(m+1)i}^{(m)}\label{eq:3.4}
\end{equation}
and
\begin{equation}
F^{(m,\mu)}(z)=\sum_{k=0}^{\infty}\frac{|z|^{2k}}{D_{k}^{(m,\mu)2}}.\label{eq:3.5}
\end{equation}
Using the form (\ref{eq:2.5}) for the $a_{\nu}^{(m)}$ gives
\begin{equation}
D_{k}^{(m,\mu)}=(-)^{k}(2m+2)^{k(m+1)/2}[(\frac{\mu-1}{m+1}+1)_{k}(\frac{\mu-2}{m+1}+1)_{k}\dots(\frac{\mu-m}{m+1}+1)_{k}(\frac{\mu+m+1}{m+1}+1)_{k}]^{\frac{1}{2}}\label{eq:3.6}
\end{equation}
and
\begin{equation}
F^{(m,\mu)}(z)=\phantom{|}_{1}F_{m+1}(1;\frac{\mu-1}{m+1}+1,\frac{\mu-2}{m+1}+1,\dots,\frac{\mu-m}{m+1}+1,\frac{\mu+m+1}{m+1}+1;\frac{|z|^{2}}{(2m+2)^{m+1}}).\label{eq:3.7}
\end{equation}
Here
\begin{equation}
(a)_{k}=\frac{\Gamma(a+k)}{\Gamma(a)}\label{eq:3.8}
\end{equation}
and $\phantom{|}_{1}F_{m+1}$ is a generalized hypergeometric function
\cite{Gradsteyn1980}.

We define a linearized version, $\tilde{c}(m),$ of the $c(m)$ operator
(in the intrinsic class of \cite{Fernandez1994}) to have matrix elements
\begin{equation}
\langle\,\nu-m-1\,(-)\,|\,\tilde{c}(m)\,|\,\nu\,(-)\,\rangle=\sqrt{2\nu},\label{eq:3.9}
\end{equation}
like those of the harmonic oscillator annihilation operator. In operator
form, this is (for $\tilde{c}$ acting on the subspace with lowest
weight $\mu$)
\begin{equation}
\tilde{c}(m,\mu)=-\frac{1}{[Q_{m}(H^{-})]^{\frac{1}{2}}}\sqrt{\frac{H^{-}-2m-2-2\mu}{m+1}}\,c.\label{eq:3.9.1}
\end{equation}
The operators $\tilde{c}$ and $\tilde{c}^{\dagger}$ satisfy the
Heisenberg algebra
\begin{align}
[H^{-},\tilde{c}^{\dagger}] & =(2m+2)\tilde{c}^{\dagger},\quad[H^{-},\tilde{c}]=-(2m+2)\tilde{c},\nonumber \\
\tilde{c}(m,\mu)\tilde{c}^{\dagger}(m,\mu) & =\frac{H^{-}-2m-2-2\mu}{m+1}+2,\quad\tilde{c}^{\dagger}(m,\mu)\tilde{c}(m,\mu)=\frac{H^{-}-2m-2-2\mu}{m+1},\nonumber \\{}
[\tilde{c}(m,\mu),\tilde{c}^{\dagger}(m,\mu)] & =2.\label{eq:3.9.2}
\end{align}

Because of the similarity with the harmonic oscillator, it is a simple
matter to find the form of the coherent states, again with $m+1$
orthogonal ladders labelled by the lowest weight $\mu$
\begin{equation}
|\,z,\tilde{c},m,\mu\,\rangle=\sum_{k=0}^{\infty}|\,\mu+(m+1)k,m\,(-)\,\rangle\tilde{A}_{k}(z),\label{eq:3.10}
\end{equation}
with
\begin{equation}
\tilde{A}_{k}(z)=e^{-|z|^{2}/4}\frac{(z/\sqrt{2})^{k}}{\sqrt{k!}}.\label{eq:3.11}
\end{equation}

\section{Energy expectations}

The energy eigenvalues are given by
\begin{equation}
H^{-}\,|\,\mu+(m+1)k,m\,\rangle=(2\mu+(2m+2)(k+1))\,|\,\mu+(m+1)k,m\,\rangle.\label{eq:4.1}
\end{equation}
This leads to the energy expectations
\begin{multline}
\langle\,z,c,m,\mu\,|\,H^{-}\,|\,z,c,m,\mu\,\rangle=2\mu+2m+2+\frac{2m+2}{(\mu+m)(\mu+m-1)\dots(\mu+1)(\mu+2m+2)}\times\\
\times\frac{|z|^{2}}{2^{m+1}}\frac{\phantom{|}_{1}F_{m+1}(2;\frac{\mu+m}{m+1}+1,\frac{\mu+m-1}{m+1}+1,\dots,\frac{\mu+1}{m+1}+1,\frac{\mu+2m+2}{m+1}+1;|z|^{2}/(2m+2)^{m+1})}{\phantom{|}_{1}F_{m+1}(1;\frac{\mu+m}{m+1},\frac{\mu+m-1}{m+1},\dots,\frac{\mu+1}{m+1},\frac{\mu+2m+2}{m+1};|z|^{2}/(2m+2)^{m+1})}.\label{eq:4.2}
\end{multline}

For the linearized operators, we have the much simpler result
\begin{equation}
\langle\,z,\tilde{c},m,\mu\,|\,H^{-}\,|\,z,\tilde{c},m,\mu\,\rangle=2\mu+2m+2+(m+1)|z|^{2}.\label{eq:4.3}
\end{equation}

In Figure 1, we show these functions of $|z|$ for $m=4$ and in Figure
2 for $m=6.$ This allows comparison between different values of $m.$
We see that the average energy always rises much faster for the $\tilde{c}$
coherent states than for the $c$ coherent states.

\begin{figure}
\noindent \begin{centering}
\includegraphics[width=14cm]{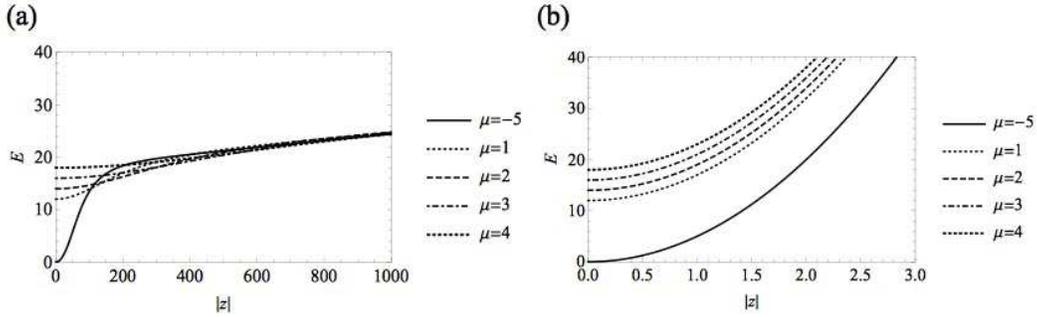}
\par\end{centering}
\caption{Energy expectations for (a) $\mathrm{CS}(c(4),\mu)$ and (b) $\mathrm{CS}(\tilde{c}(4),\mu),$
for $\mu=-5,1,2,3,4.$}

\end{figure}

\begin{figure}
\noindent \begin{centering}
\includegraphics[width=14cm]{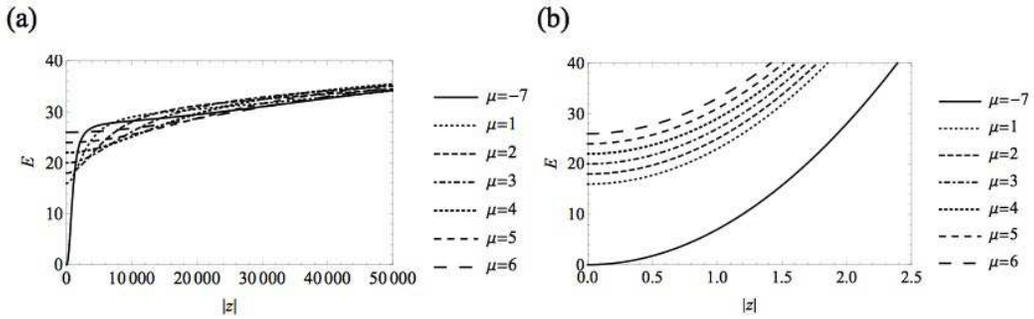}
\par\end{centering}
\caption{Energy expectations for (a) $\mathrm{CS}(c(6),\mu)$ and (b) $\mathrm{CS}(\tilde{c}(6),\mu),$
for $\mu=-7,1,2,3,4,5,6.$}

\end{figure}

\section{Time evolution of position probability densities}

We study the time evolution of the position probability densities
for the coherent states we have constructed. For the $\mathrm{CS}(c(m),\mu)$
states, we find that for low values of $|z|$ the pattern of many
peaks and valleys does not take a simple form. However, in all cases,
we find that increasing $|z|$ to very large values gives a pattern
of a small number of peaks oscillating in the potential and colliding
with each other, producing interference fringes when they collide.
We call this regime semi-classical, since it is the closest to the
appearance of (more than one) classical particle that we have found.

In our previous paper \cite{Hoffmann2018a}, we only studied the $\mathrm{CS}(c(2),\mu)$
coherent states to $z=100,$ and did not see the semi-classical regime.
Taking $z=17\,000$ for $\mathrm{CS}(c(2),-3)$ (to give the same
average energy as $\mathrm{CS}(\tilde{c}(2),-3)$ with $z=15$) shows
the semi-classical regime with three wavepackets.

It is now clear that $\mathrm{CS}(c(m),\mu)$ and $\mathrm{CS}(\tilde{c}(m),\mu)$
coherent states show very similar behaviour when compared at the same
average energy, but not at the same value of the parameter $z.$

For general $\mathrm{CS}(c(m),\mu),$ the position probability density
is given by
\begin{align}
\rho(x,t;z,m,\mu) & =|\langle\,x\,|\,e^{-iH^{-}t}\,|\,z,c,m,\mu\,\rangle|^{2}\nonumber \\
 & =|\sum_{k=0}^{\infty}\psi_{\mu+(m+1)k}^{(m)}(x)A_{k}^{(m,\mu)}(z\,e^{-i(2m+2)t})|^{2}.\label{eq:5.1}
\end{align}
For the linearized states $\mathrm{CS}(\tilde{c}(m),\mu)$, the corresponding
expression is
\begin{equation}
\tilde{\rho}(x,t;z,m,\mu)=|\sum_{k=0}^{\infty}\psi_{\mu+(m+1)k}^{(m)}(x)\tilde{A}_{k}(z\,e^{-i(2m+2)t})|^{2}.\label{eq:5.2}
\end{equation}

We find that the coefficients $A_{k}^{(m,\mu)}(z)$ fall off very
rapidly with $k$ for the values of $z$ we are using. So in practice,
we only take the sum in (\ref{eq:5.1}) to a small cutoff value of
$k.$

The state vectors $|\,z,c,m,\mu\,\rangle$ and $|\,\tilde{z},\tilde{c},m,\mu\,\rangle$
have the same average energy and similar behaviour for $|\tilde{z}|$
much less than $|z|,$ as can be seen in Figures 1 and 2. For the
$\mathrm{CS}(\tilde{c}(m),\mu)$ states at low $|\tilde{z}|,$ we
can also cut off the sums over $k$ at low values.

For $\mathrm{CS}(c(4),-5)$ at $z=100\,000$ and $\mathrm{CS}(\tilde{c}(4),-5)$
at $z=4.7$ (giving the same average energy $\langle\,E\,\rangle=108$)
we show the probability densities in Figure 3 and note similar behaviour,
both systems being in the semi-classical regime. As noted above, in
this regime we see isolated wavepackets oscillating and colliding.

For $m=6$, we display the time-dependence of the position probability
density at a set of discrete times. These plots, unlike the 3D plots,
do not smear the interference fringes. We note that at times when
the wavepackets are well-separated, we count $m+1$ wavepackets, at
least for these two choices of $m$ and recalling the result for $m=2$.
For $m=6,$ we started with low values of (real, positive) $z$ and
found indistinct position probability densities. It was necessary
to take $z$ to the extremely large value of $10^{8}$ before the
pattern shown below emerged.

\begin{figure}
\noindent \begin{centering}
\includegraphics[width=12cm]{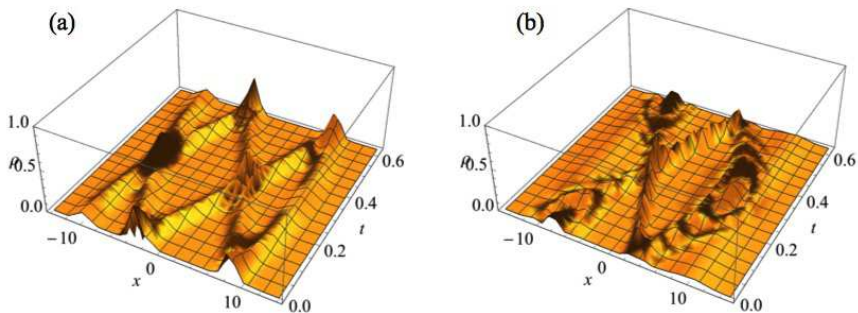}
\par\end{centering}
\caption{Densities for (a) $\mathrm{CS}(c(4),-5)$ at $z=100\,000$ and (b)
$\mathrm{CS}(\tilde{c}(4),-5)$ at $z=4.7.$}

\end{figure}

\begin{figure}
\noindent \begin{centering}
\includegraphics[width=12cm]{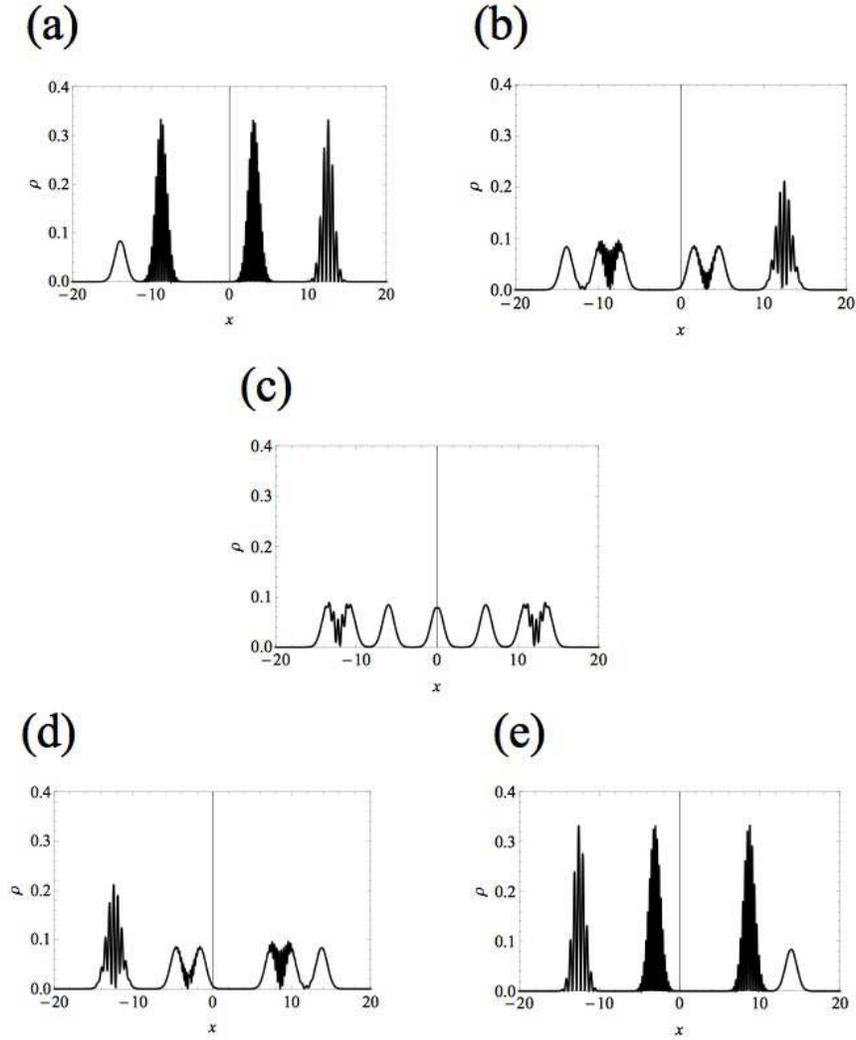}
\par\end{centering}
\caption{Density for $\mathrm{CS}(c(6),-7)$ at $z=10^{8}$ for (a) $t=0,$
(b) $t=T/8,$ (c) $t=T/4,$ (d) $t=3T/8$ and (e) $t=T/2$ with period
$T=\pi/7.$ }

\end{figure}

\section{Schrödinger cat states}

A cat state is a superposition of two macroscopically distinguishable
states. We construct even (+) and odd (-) states by (for $z$ real)
\begin{equation}
|\,z,c,m,\mu,\pm\,\rangle=\frac{1}{\sqrt{2}}\{|\,+z,c,m,\mu\,\rangle\pm|\,-z,c,m,\mu\,\rangle\}.\label{eq:6.1}
\end{equation}
A measure of the distinguishability of the two component states is
(for $m=6$)
\begin{equation}
D(|z|,\mu)=\langle\,+z,c,6,\mu,+\,|\,-z,c,6,\mu\,\rangle=\frac{\phantom{|}_{1}F_{7}(1;\frac{\mu+6}{7},\frac{\mu+5}{7},\dots,\frac{\mu+1}{7},\frac{\mu+14}{7};-|z|^{2}/14^{7})}{\phantom{|}_{1}F_{7}(1;\frac{\mu+6}{7},\frac{\mu+5}{7},\dots,\frac{\mu+1}{7},\frac{\mu+14}{7};|z|^{2}/14^{7})}.\label{eq:6.2}
\end{equation}
This is shown in Figure 5, where we see that $z=10^{8}$ is certainly
sufficient for distinguishability.

\begin{figure}
\noindent \begin{centering}
\includegraphics[width=10cm]{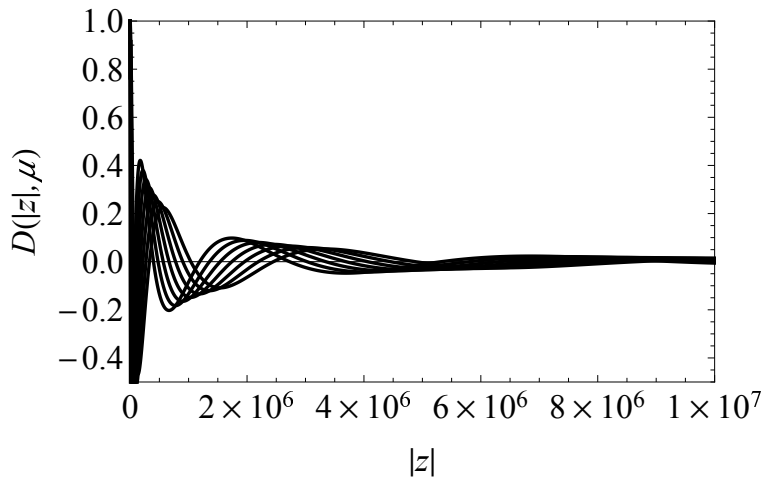}
\par\end{centering}
\caption{Distinguishability measures for $m=6$ coherent states $\mathrm{CS}(c(6),\mu),$
$\mu=-7,1,2,3,4,5,6.$}

\end{figure}

The time-dependent position probability densities for these two cat
states are shown in a density plot in Figure 6. We see a great deal
of structure at this very large value of $z.$ The odd state has a
central nodal line, consistent with the parity of the odd wavefunctions.

\begin{figure}
\noindent \begin{centering}
\includegraphics[width=12cm]{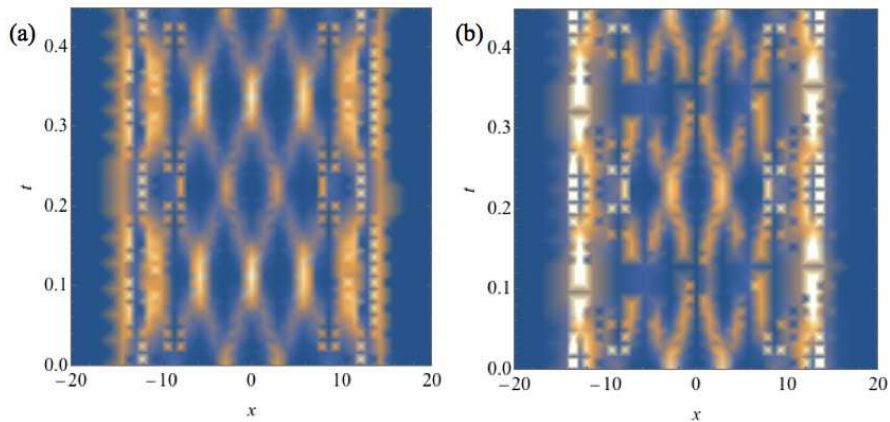}
\par\end{centering}
\caption{Time-dependent probability densities for (a) even and (b) odd cat
states, for $c(6),$ $\mu=-7.$}

\end{figure}

\section{Wigner functions}

The Wigner function for a single particle is an everywhere real function
of position, $x,$ and momentum, $p,$ so can be considered a distribution
in phase space. For the coherent states of the harmonic oscillator,
the Wigner functions are everywhere positive Gaussians centred on
$x=\mathrm{Re}\,z$ and $p=\mathrm{Im}\,z.$ This fact is used to
reinforce the assertion, that we arrived at by looking at the evolution
of the position probability density, that the coherent states of the
harmonic oscillator are the most closely classical states of a one-dimensional
system. It is therefore judged that a state vector does not behave
classically, does not have a classical analogue, if its Wigner function
is anywhere negative. We will use this criterion to judge the coherent
states of our $c(m)$ operators.

For a general state vector $|\,\psi\,\rangle,$ the Wigner function
is defined as
\begin{equation}
W(x,p;\psi)=\frac{1}{\pi}\int_{-\infty}^{\infty}dy\,\langle\,\psi\,|\,x-y\,\rangle\langle\,x+y\,|\,\psi\,\rangle e^{-i2py}.\label{eq:7.1}
\end{equation}
For $|\,\psi\,\rangle=|\,z,c,m,\mu\,\rangle,$we calculate the Wigner
function using
\begin{equation}
W(x,p:z,c,m,\mu)=\sum_{k_{1}=0}^{\infty}\sum_{k_{2}=0}^{\infty}A_{k_{1}}^{(m,\mu)*}(z)w_{k_{1}k_{2}}^{(m,\mu)}(x,p)A_{k_{2}}^{(m,\mu)}(z),\label{eq:7.2}
\end{equation}
with
\begin{equation}
w_{k_{1}k_{2}}^{(m,\mu)}(x,p)=\frac{1}{\pi}\int_{-\infty}^{\infty}dy\,\psi_{\mu+(m+1)k_{1}}^{(m)*}(x-y)\psi_{\mu+(m+1)k_{2}}^{(m)}(x+y)e^{-i2py}.\label{eq:7.3}
\end{equation}
As we noted in Section V, the coefficients $A_{k}^{(m,\mu)}(z)$ fall
off rapidly with $k$ for the small values of $z$ that we consider.
We found that it was sufficient to take the sums over $k_{1},k_{2}$
to $k_{1}=k_{2}=10.$

For $m=6$ we found the seven Wigner functions shown in Figure 7.
The flat, monotone areas in each graph show where the Wigner function
takes on negative values. We conclude that these coherent states do
not conform to classical expectations (except for $\mathrm{CS}(c(6),-7)$),
as do the harmonic oscillator coherent states.

\begin{figure}
\noindent \begin{centering}
\includegraphics[width=14cm]{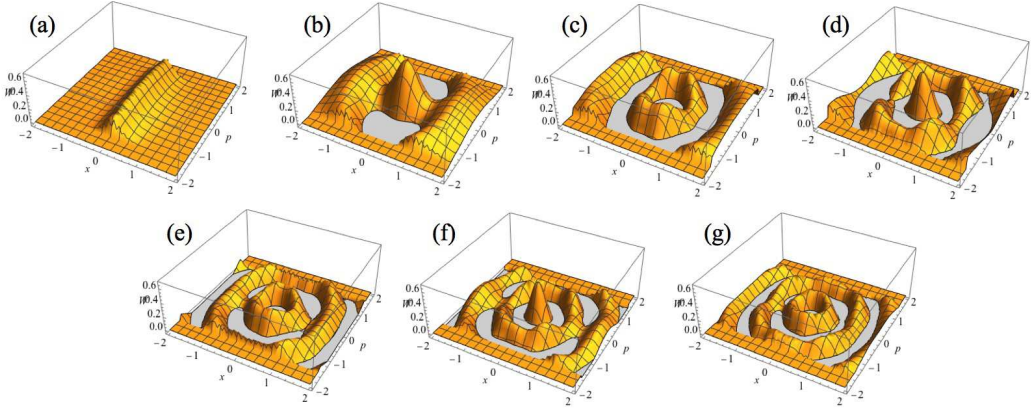}
\par\end{centering}
\caption{Wigner functions for $\mathrm{CS}(c(6),\mu)$ and $z=10,$ for $\mu$
values (a) -7, (b) 1, (c) 2, (d) 3, (e) 4, (f) 5, (g) 6.}

\end{figure}

\section{Heisenberg uncertainty relations and squeezing}

For our various coherent states, $c(m,\mu)$ and $\tilde{c}(m,\mu),$
we calculate the standard deviations in position and momentum using
\begin{align}
\sigma_{x} & =\sqrt{\langle\,x^{2}\,\rangle-\langle\,x\,\rangle^{2}},\nonumber \\
\sigma_{p} & =\sqrt{\langle\,p^{2}\,\rangle-\langle\,p\,\rangle^{2}},\label{eq:8.1}
\end{align}
where the expectation values are in the particular state considered.
Then we form the products
\begin{equation}
h(z,m,\mu)=\sigma_{x}(z,m,\mu)\sigma_{p}(z,m,\mu).\label{eq:8.2}
\end{equation}
These are required by Heisenberg's uncertainty principle to be, in
all cases, greater than or equal to $1/2.$

The expectation are calculated using, with $|\,\psi\,\rangle=|\,z,c,m,\mu\,\rangle$
again,
\begin{eqnarray}
\langle\,\psi\,|\,\hat{x}\,|\,\psi\,\rangle & = & \sum_{k_{1}}\sum_{k_{2}}A_{k_{1}}^{(m,\mu)*}(z)M_{k_{1}k_{2}}^{(x)}A_{k_{2}}^{(m,\mu)}(z),\nonumber \\
\langle\,\psi\,|\,\hat{x}^{2}\,|\,\psi\,\rangle & = & \sum_{k_{1}}\sum_{k_{2}}A_{k_{1}}^{(m,\mu)*}(z)M_{k_{1}k_{2}}^{(x^{2})}A_{k_{2}}^{(m,\mu)}(z),\nonumber \\
\langle\,\psi\,|\,\hat{p}\,|\,\psi\,\rangle & = & \sum_{k_{1}}\sum_{k_{2}}A_{k_{1}}^{(m,\mu)*}(z)M_{k_{1}k_{2}}^{(p)}A_{k_{2}}^{(m,\mu)}(z),\nonumber \\
\langle\,\psi\,|\,\hat{p}^{2}\,|\,\psi\,\rangle & = & \sum_{k_{1}}\sum_{k_{2}}A_{k_{1}}^{(m,\mu)*}(z)M_{k_{1}k_{2}}^{(p^{2})}A_{k_{2}}^{(m,\mu)}(z),\label{eq:8.3}
\end{eqnarray}
with
\begin{eqnarray}
M_{k_{1}k_{2}}^{(x)} & = & \int_{-\infty}^{\infty}dx\,\psi_{\mu+(m+1)k_{1}}^{(m)*}(x)\,x\,\psi_{\mu+(m+1)k_{2}}^{(m)}(x),\nonumber \\
M_{k_{1}k_{2}}^{(x^{2})} & = & \int_{-\infty}^{\infty}dx\,\psi_{\mu+(m+1)k_{1}}^{(m)*}(x)\,x^{2}\psi_{\mu+(m+1)k_{2}}^{(m)}(x),\nonumber \\
M_{k_{1}k_{2}}^{(p)} & = & \int_{-\infty}^{\infty}dx\,\psi_{\mu+(m+1)k_{1}}^{(m)*}(x)(-i\frac{d}{dx})\psi_{\mu+(m+1)k_{2}}^{(m)}(x),\nonumber \\
M_{k_{1}k_{2}}^{(p^{2})} & = & \int_{-\infty}^{\infty}dx\,\psi_{\mu+(m+1)k_{1}}^{(m)*}(x)(-\frac{d^{2}}{dx^{2}})\psi_{\mu+(m+1)k_{2}}^{(m)}(x).\label{eq:8.4}
\end{eqnarray}
Again the sums over $k_{1},k_{2}$ can be taken to a small finite
value depending on the range of $z$ values being considered.

In every case we considered, we found a product $h(z,m,\mu)$ greater
than $1/2,$ as expected. In Figure 8 we show the product as a function
of $\mathrm{Re}\,z$ and $\mathrm{Im}\,z$ for $\mathrm{CS}(c(4),-5)$
and for $\mathrm{CS}(\tilde{c}(6),-7).$

\begin{figure}
\noindent \begin{centering}
\includegraphics[width=12cm]{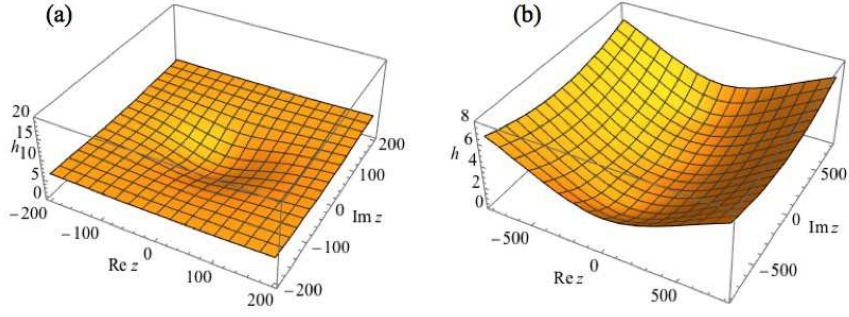}
\par\end{centering}
\caption{Heisenberg uncertainty relations for (a) $\mathrm{CS}(c(4),-5)$ and
(b) $\mathrm{CS}(\tilde{c}(6),-7).$}

\end{figure}

The Heisenberg uncertainty principle for one-dimensional systems with
position and momentum operators, as we have been considering, is $\sigma_{x}\sigma_{p}\geq1/2.$
This bound is saturated for the coherent states of the harmonic oscillator,
where $\sigma_{x}=\sigma_{p}=1/\sqrt{2}$ in our units. It is possible
for one of the standard deviations to be less that the harmonic oscillator
value, $1/\sqrt{2},$ provided that the other standard deviation is
greater than $1/\sqrt{2},$ to preserve the uncertainty relation.
In these cases we say that we are dealing with a squeezed state \cite{Loudon1987,Walls1983}.

We test our coherent states for squeezing and find that it occurs
in many cases. For example, Figure 9 shows the observation of squeezing
for the $\mathrm{CS}(\tilde{c}(4),-5)$ state in $\sigma_{x}$ at
$t=0.$ The graph shows a flat area for $|z|\apprle0.7$ where $\sigma_{x}$
falls below $1/\sqrt{2}.$

Squeezing has physical applications. Light can be squeezed and used
to reduce the noise in measurements \cite{Oelker2016}.

\begin{figure}
\noindent \begin{centering}
\includegraphics[width=8cm]{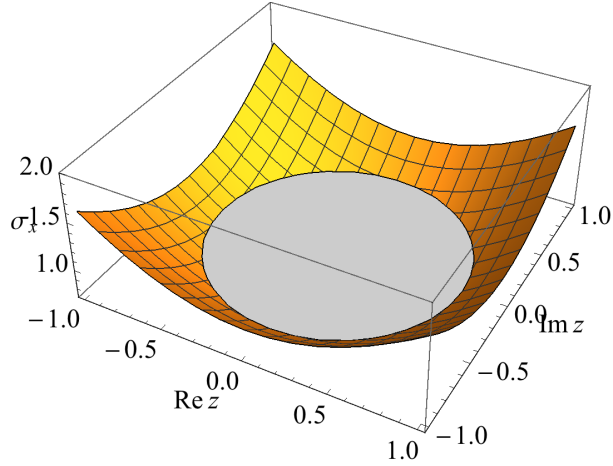}
\par\end{centering}
\caption{Observation of squeezing for $\mathrm{CS}(\tilde{c}(4),-5)$ (plot
of the position standard deviation, $\sigma_{x},$ as a function of
$\mathrm{Re}\,z$ and $\mathrm{Im}\,z$).}

\end{figure}

\section{Number statistics}

The coherent states of the harmonic oscillator have an excitation
number probability distribution (from (\ref{eq:3.11}))
\begin{equation}
P_{n}(z)=|\langle\,n\,|\,z\,\rangle|^{2}=e^{-|z|^{2}/2}\frac{(|z|^{2}/2)^{n}}{n!},\label{eq:10.1}
\end{equation}
which is a Poisson distribution with mean number
\begin{equation}
\langle\,N\,\rangle=\frac{|z|^{2}}{2}.\label{eq:10.2}
\end{equation}
It is easily seen that the number variance satisfies
\begin{equation}
\Delta N^{2}=\langle\,N^{2}\,\rangle-\langle\,N\,\rangle^{2}=\langle\,N\,\rangle.\label{eq:10.3}
\end{equation}

Then a number distribution with variance less than the average number
is called sub-Poissonian, and called super-Poissonian if the variance
is greater than the mean. The Mandel $Q$ parameter measures the classification
of an arbitrary number distribution:
\begin{equation}
Q=\frac{\Delta N^{2}-\langle\,N\,\rangle}{\langle\,N\,\rangle}=\frac{\langle\,N(N-1)\,\rangle-\langle\,N\,\rangle^{2}}{\langle\,N\,\rangle}.\label{eq:10.4}
\end{equation}
It vanishes for a Poisson distribution, is negative for a sub-Poissonian
distribution and is positive for a super-Poissonian distribution.

We investigate the number statistics of the coherent states labelled
$\mathrm{CS}(c(4),\mu)$. We define a number operator, $N_{\mu}$,
that acts on the subspace of the order $4$ system with lowest weight
$\mu$ by
\begin{equation}
N_{\mu}\,|\,\mu+5k,4\,(-)\,\rangle=k\,|\,\mu+5k,4\,(-)\,\rangle\label{eq:10.6}
\end{equation}
Then we calculate the Mandel parameter as $Q(c(4),\mu,z)$ where the
expectations in (\ref{eq:10.4}) are in the state vector $|\,z,c,4,\mu\,\rangle$
and $N=N_{\mu}.$

We find
\begin{equation}
\langle\,N_{\mu}\,\rangle=\frac{1}{(\mu+4)(\mu+3)(\mu+2)(\mu+1)(\mu+10)}\frac{|z|^{2}}{32}\frac{\phantom{|}_{1}F_{5}(2;\frac{\mu+9}{5},\frac{\mu+8}{5},\frac{\mu+7}{5},\frac{\mu+6}{5},\frac{\mu+15}{5};|z|^{2}/10^{5})}{\phantom{|}_{1}F_{5}(1;\frac{\mu+4}{5},\frac{\mu+3}{5},\frac{\mu+2}{5},\frac{\mu+1}{5},\frac{\mu+10}{5};|z|^{2}/10^{5})}\label{eq:10.7}
\end{equation}
and
\begin{equation}
\langle\,N_{\mu}(N_{\mu}-1)\,\rangle=\frac{|z|^{4}}{2^{9}}\frac{\phantom{|}_{1}F_{5}(3;\frac{\mu+14}{5},\frac{\mu+13}{5},\frac{\mu+12}{5},\frac{\mu+11}{5},\frac{\mu+20}{5};|z|^{2}/10^{5})/\phantom{|}_{1}F_{5}(1;\frac{\mu+4}{5},\frac{\mu+3}{5},\frac{\mu+2}{5},\frac{\mu+1}{5},\frac{\mu+10}{5};|z|^{2}/10^{5})}{(\mu+4)(\mu+3)(\mu+2)(\mu+1)(\mu+10)(\mu+9)(\mu+8)(\mu+7)(\mu+6)(\mu+15)}.\label{eq:10.8}
\end{equation}
This gives the Mandel $Q$ parameter as a function of $|z|$ for $\mu=-5$
as shown in Figure 10. We see that the parameter is negative except
at the origin, indicating sub-Poissonian statistics. This is considered
another sign of non-classical behaviour. For other values $\mu=1,2,3,4,$
curves of similar appearance were produced, but with progressively
slower variation as $\mu$ was increased.

\begin{figure}
\begin{centering}
\includegraphics[width=10cm]{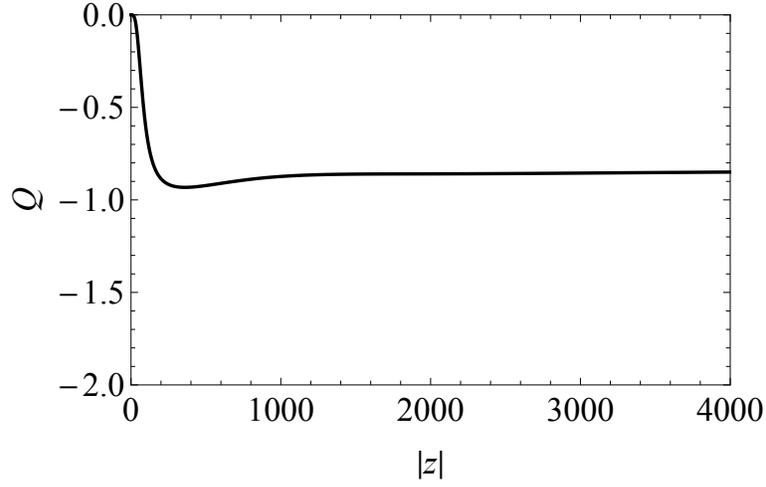}
\par\end{centering}
\caption{Mandel $Q$ parameter as a function of $|z|$ for the coherent states
$\mathrm{CS}(c(4),-5).$}

\end{figure}

It should be clear that since the structure of the linearized $\mathrm{CS}(\tilde{c}(m),\mu)$
coherent states is so like that of the harmonic oscillator coherent
states, they have only Poisson statistics. To verify this, we find
\begin{equation}
\langle\,N_{\mu}\,\rangle=\sum_{k=0}^{\infty}k\,|\tilde{A}_{k}(z)|^{2}=\sum_{k=0}^{\infty}k\,e^{-|z|^{2}/2}\frac{(|z|^{2}/2)^{k}}{k!}=\frac{|z|^{2}}{2}\label{eq:10.9}
\end{equation}
and
\begin{equation}
\langle\,N_{\mu}(N_{\mu}-1)\,\rangle=\sum_{k=0}^{\infty}k(k-1)\,|\tilde{A}_{k}(z)|^{2}=\sum_{k=0}^{\infty}k(k-1)\,e^{-|z|^{2}/2}\frac{(|z|^{2}/2)^{k}}{k!}=\left(\frac{|z|^{2}}{2}\right)^{2}.\label{eq:10.10}
\end{equation}
So
\begin{equation}
Q(\tilde{c}(m),\mu,z)\equiv0\quad\mathrm{for}\ \mathrm{all}\ m,\mu.\label{eq:10.11}
\end{equation}

\section{A coherent state on a beamsplitter}

The optical device called a beamsplitter is, in its simplest form,
a partially silvered mirror \cite{Walls2008}. For a 50:50 beamsplitter
(the only case we will consider) half the intensity of a beam of light
incident on the beamsplitter is transmitted while the other half is
reflected. If two perpendicular beams are incident on the beamsplitter,
they are mixed coherently and two perpendicular output beams are produced.
If just one incident beam is used, the input in the other direction
is the vacuum.

We can place one of our coherent states on one arm of a beamsplitter,
with the excitations above the ground state taking the role of photons.
We confirm below that if the input is a harmonic oscillator coherent
state, representing coherent light, the ``photon'' number distribution
from the two output arms is uncorrelated, factorizing into the product
of two independent Poisson number distributions. If a squeezed state
is the input, there is correlation and entanglement between the two
output arms \cite{Sanders2012,Ourjoumtsev2009}. Entangled states
may have applications in quantum cryptography and quantum teleportation
\cite{Jennewein2000,Bennett1993}. Thus a beamsplitter calculation
can give us more information on whether a coherent state displays
classical behaviour (like the harmonic oscillator coherent states)
or non-classical behaviour.

If $\tilde{c}_{1},\tilde{c}_{1}^{\dagger}$ and $\tilde{c}_{2,}\tilde{c}_{2}^{\dagger}$
are linearized ladder operators for the two input arms (for indices
$m$ and $\mu$), respectively, with harmonic oscillator commutation
relations, the action of the beamsplitter is represented by the unitary
transformation \cite{Kim2002}
\begin{equation}
U=\exp(i\frac{\pi}{8}(\tilde{c}_{1}\tilde{c}_{2}^{\dagger}+\tilde{c}_{2}\tilde{c}_{1}^{\dagger})),\label{eq:11.1}
\end{equation}
which is seen to give the transformations
\begin{equation}
U\tilde{c}_{1}^{\dagger}U^{\dagger}=\frac{1}{\sqrt{2}}(\tilde{c}_{1}^{\dagger}+i\tilde{c}_{2}^{\dagger}),\quad U\tilde{c}_{2}^{\dagger}U^{\dagger}=\frac{1}{\sqrt{2}}(i\tilde{c}_{1}^{\dagger}+\tilde{c}_{2}^{\dagger}).\label{eq:11.2}
\end{equation}
The factors of $i$ represent $\pi/2$ phase shifts upon reflection.
Then it can be shown that
\begin{equation}
U\{|\,n\,\rangle\otimes|\,0\,\rangle\}=\frac{1}{2^{\frac{n}{2}}}\sum_{r=0}^{n}\binom{n}{r}^{\frac{1}{2}}i^{r}|\,n-r\,\rangle\otimes|\,r\,\rangle,\label{eq:11.3}
\end{equation}
where
\begin{equation}
|\,n\,\rangle=|\,\mu+(m+1)n,m\,\rangle.\label{eq:11.4}
\end{equation}
For a general coherent state with coefficients $A_{n}(z),$
\begin{equation}
|\,z\,\rangle=\sum_{n=0}^{\infty}A_{n}(z)\,|\,n\,\rangle,\label{eq:11.5}
\end{equation}
we have the result for the output state
\begin{equation}
|\,\mathrm{out}\,\rangle=U\{|\,z\,\rangle\otimes|\,0\,\rangle\}=\sum_{n=0}^{\infty}\frac{A_{n}(z)}{2^{\frac{n}{2}}}\sum_{r=0}^{n}\binom{n}{r}^{\frac{1}{2}}i^{r}|\,n-r\,\rangle\otimes|\,r\,\rangle.\label{eq:11.6}
\end{equation}
Then the two-photon-number probability distribution for the output
arms is
\begin{align}
P(n_{1},n_{2};z) & =|\{\langle\,n_{1}\,|\otimes\langle\,n_{2}\,|\}\,U\{|\,z\,\rangle\otimes|\,0\,\rangle\}|^{2}=\frac{|A_{n_{1}+n_{2}}(z)|^{2}}{2^{n_{1}+n_{2}}}\binom{n_{1}+n_{2}}{n_{2}}.\label{eq:11.7}
\end{align}

We check for a harmonic oscillator coherent state input,
\begin{equation}
A_{k}^{(0)}(z)=e^{-|z|^{2}/4}(\frac{z}{\sqrt{2}})^{k}\frac{1}{\sqrt{k!}},\label{eq:11.8}
\end{equation}
that
\begin{equation}
P^{(0)}(n_{1},n_{2};z)=p_{n_{1}}(\frac{|z|^{2}}{4})p_{n_{2}}(\frac{|z|^{2}}{4}),\label{eq:11.9}
\end{equation}
where
\begin{equation}
p_{n}(\bar{n})=\frac{e^{-\bar{n}}\bar{n}^{n}}{n!}\label{eq:11.10}
\end{equation}
is the Poisson distribution for mean number $\bar{n}.$ The mean numbers
in each arm are half the value for the input state.

For the case $\mathrm{CS}(c(4),-5),$ we found
\begin{multline}
P^{(4,-5)}(n_{1},n_{2};z)=\frac{1}{n_{1}!n_{2}!}\frac{1}{F^{(4,-5)}(z)}\left(\frac{|z|^{2}}{2\times10^{5}}\right)^{n_{1}+n_{2}}\times\\
\times\frac{\Gamma(-\frac{1}{5})\Gamma(-\frac{2}{5})\Gamma(-\frac{3}{5})\Gamma(-\frac{4}{5})}{\Gamma(n_{1}+n_{2}-\frac{1}{5})\Gamma(n_{1}+n_{2}-\frac{2}{5})\Gamma(n_{1}+n_{2}-\frac{3}{5})\Gamma(n_{1}+n_{2}-\frac{4}{5})}.\label{eq:11.11}
\end{multline}
The presence of functions of $n_{1}+n_{2}$ shows that this probability
distribution does not take a factorized form, indicating correlations
between the two output arms. We plot this distribution for $z=10^{5}$
and compare this with the harmonic oscillator case for $z=3.5,$ in
Figure 11.

\begin{figure}
\noindent \begin{centering}
\includegraphics[width=14cm]{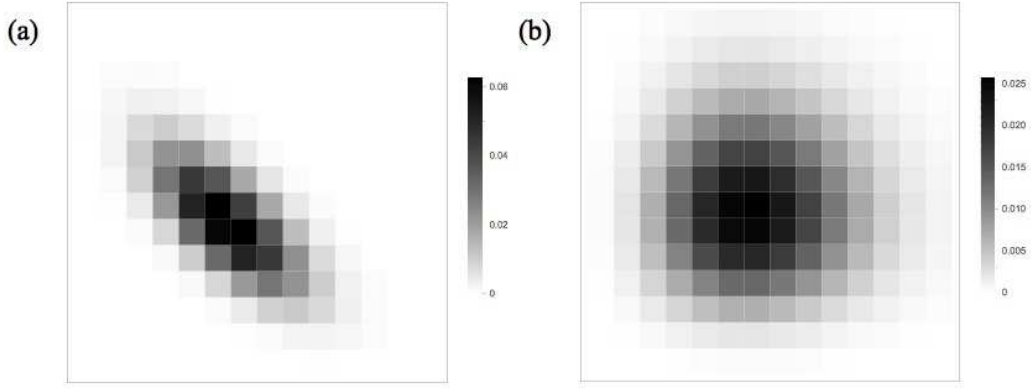}
\par\end{centering}
\caption{Two-photon-number probability density for (a) the case $\mathrm{CS}(c(4),-5)$
with $z=10^{5}$ and (b) the harmonic oscillator case with $z=3.5$. }

\end{figure}

It is evident that using the coherent states $\mathrm{CS}(\tilde{c}(m),\mu)$
would give the same result as the harmonic oscillator case.

The linear entropy is widely used as an approximate measure of entanglement
in quantum systems, defined by \cite{Affleck2009,Gerry2009}
\begin{equation}
S=1-\mathrm{Tr}\,(\,\rho_{a}^{2}),\label{eq:11.12}
\end{equation}
where the density matrix in our case is (using Eq. (\ref{eq:11.6}))
\begin{equation}
\rho_{ab}=|\,\mathrm{out}\,\rangle\langle\,\mathrm{out}\,|\label{eq:11.13}
\end{equation}
and
\begin{equation}
\rho_{a}=\sum_{r=0}^{\infty}\langle\,r\,|\,\rho_{ab}\,|\,r\,\rangle\label{eq:11.14}
\end{equation}
is the partial trace of the full density matrix over the quantum number
of the second of the systems in the bipartite direct product. The
linear entropy can take values in the range $0\leq S<1.$

We find the result for general coherent state coefficients $A_{n}(z)$,
\begin{equation}
S(z)=1-\sum_{r_{1}=0}^{\infty}\sum_{r_{2}=0}^{\infty}\left|\sum_{\kappa=0}^{\infty}G(\kappa+r_{1},r_{1};z)G^{*}(\kappa+r_{2},r_{2};z)\right|^{2},\label{eq:11.15}
\end{equation}
with
\begin{equation}
G(k,r;z)=A_{k}(z)\frac{1}{2^{\frac{k}{2}}}\binom{k}{r}^{\frac{1}{2}}.\label{eq:11.16}
\end{equation}

For a harmonic oscillator coherent state on one input arm of a beamsplitter,
the linear entropy of the output state vanishes exactly, consistent
with the lack of correlation between the two output arms.

We calculated the linear entropy for the output state of a beamsplitter
on which is placed the vacuum and the coherent state labelled $\mathrm{CS}(c(4),-5)$
(the -5 level is the ground state), with coefficients
\begin{equation}
A_{k}^{(4,-5)}(z)=\frac{(-z/10^{\frac{5}{2}})^{k}}{\sqrt{F^{(4,-5)}(z)}}\sqrt{\frac{\Gamma(-\frac{1}{5})\Gamma(-\frac{2}{5})\Gamma(-\frac{3}{5})\Gamma(-\frac{4}{5})}{\Gamma(k-\frac{1}{5})\Gamma(k-\frac{2}{5})\Gamma(k-\frac{3}{5})\Gamma(k-\frac{4}{5})k!}}.\label{eq:11.17}
\end{equation}
in the superposition
\begin{equation}
|\,z,c,4,-5\,\rangle=\sum_{k=0}^{\infty}|\,-5+5k,4\,(-)\,\rangle A_{k}^{(4,-5)}(z),\label{eq:11.18}
\end{equation}
where
\begin{equation}
F^{(4,-5)}(z)=\phantom{|}_{0}F_{4}(-\frac{1}{5},-\frac{2}{5},-\frac{3}{5},-\frac{4}{5};|z|^{2}/10^{5})\label{eq:11.19}
\end{equation}
is a generalized hypergeometric function \cite{Gradsteyn1980}.
\noindent \begin{center}
\begin{figure}
\noindent \begin{centering}
\includegraphics[width=10cm]{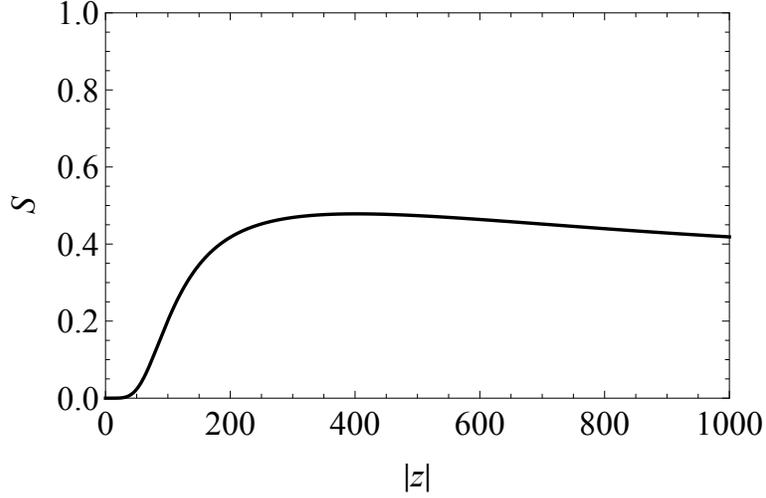}
\par\end{centering}
\caption{Linear entropy for the outgoing beamsplitter state for $\mathrm{CS}(c(4),-5)$
as a function of $|z|.$}
\end{figure}
\par\end{center}

The results are shown in Figure 12. We see a significantly nonzero
linear entropy, indicating entanglement between the two output arms
of the beamsplitter. This non-classicality is consistent with what
we found from the other measures.

\section{Conclusions}

We have constructed the $m+1$ orthogonal coherent states, using the
Barut-Girardello definition \cite{Barut1971}, for the annihilation
operator $c(m)$ (obtained by Marquette and Quesne \cite{Marquette2013b,Marquette2013a}),
for general $m.$ For comparison, we followed the same procedure for
the linearized operators $\tilde{c}(m).$ We calculated the properties
of these coherent states, plotting energy expectations, time-dependent
position probability densities for the coherent states and for the
even and odd cat states, Wigner functions, and Heisenberg uncertainty
products. We also plotted the position standard deviations to investigate
squeezing and the Mandel $Q$ parameter to investigate number statistics.
We placed one of our coherent states on one arm of a beamsplitter
and calculated the two ``photon'' number probability distribution
for the output arms. Then we calculated the linear entropy for an
example output state.

We found that the position probability density for a $c(m)$ coherent
state is rather indistinct for low values of $|z|,$ but for sufficiently
large values it separates into $m+1$ distinct wavepackets, oscillating
in the potential and colliding to produce interference fringes. We
saw the same behaviour for the $\tilde{c}(m)$ coherent states, at
much lower values of $|z|,$ but at comparable average energies. Consequently,
the even and odd cat states display a great deal of wavepacket structure.

We found the Wigner function to take on negative values for all but
the case $\mu=-7$ of $m=6,$ an indication of non-classical behaviour.
The Heisenberg uncertainty products for two cases showed minima at
$z=0$ with minimum values greater than $1/2.$ By plotting the standard
deviation in position for our coherent states as a function of $\mathrm{Re}\,z$
and $\mathrm{Im}\,z$, and finding where it fell below $1/\sqrt{2},$
we were able to observe squeezing in many of our cases.

The Mandel $Q$ parameter was found to be negative in all cases of
$m=4$ (except at $z=0,$ where it vanished) indicating sub-Poissonian
statistics and, again, non-classical behaviour.

We found correlations in the output arms two-number distribution (it
did not factorize) for one of our coherent states on a beamsplitter,
another indication of non-classical behaviour. The linear entropy
was calculated for the beamsplitter output state and was found to
take significant nonzero values (except for $z=0$), indicating the
presence of entanglement, another non-classical feature.
\begin{acknowledgments}
IM was supported by Australian Research Council Discovery Project
DP 160101376. YZZ was supported by National Natural Science Foundation
of China (Grant No. 11775177). VH acknowledges the support of research
grants from NSERC of Canada. SH receives financial support from a
UQ Research Scholarship.
\end{acknowledgments}

\bibliographystyle{vancouver}

\end{document}